\documentclass[a4paper,fleqn,colorlinks=true,linkcolor=blue,citecolor=blue,urlcolor=blue]{cas-dc}
\usepackage{amsmath}
\usepackage[numbers,sort&compress]{natbib}
\usepackage{float}

\graphicspath{{./}}

\usepackage{xcolor}
\usepackage[nameinlink,capitalise,noabbrev]{cleveref}
\crefname{figure}{Fig.}{Figs.}
\Crefname{figure}{Figure}{Figures}
\crefname{equation}{Eq.}{Eqs.}
\Crefname{equation}{Equation}{Equations}

\begin{document}
\let\WriteBookmarks\relax
\def\floatpagepagefraction{1}
\def\textpagefraction{.001}

\shorttitle{Helical pipes homogenize particle transport in pneumatic conveying}

\shortauthors{W. T. Ku et al.}  

\title [mode = title]{Helical pipes homogenize particle transport in pneumatic conveying}
\author[1]{Wing To Ku}
\ead{wing.ku@fau.de}
\credit{Software, investigation, validation, visualization, writing – first draft}
\affiliation[1]{organization={Institute for Multiscale Simulation, Friedrich-Alexander-Universit\"at Erlangen-N\"urnberg},
    addressline={Cauerstra{\ss}e 3}, 
    city={Erlangen},
    citysep={}, 
    postcode={91058}, 
    country={Germany}
    }

\author[1]{Patric M\"uller}[orcid=0000-0002-0090-4486]
\ead{patric.mueller@fau.de}
\credit{methods, Project administration, funding acquisition, writing - review \& editing}

\author[1]{Thorsten P\"oschel}[orcid=0000-0001-5913-1070]
\cormark[1]
\ead{thorsten.poeschel@fau.de}
\cortext[cor1]{Corresponding author}
\credit{conceptualization, methods, writing - review \& editing, supervision}

\begin{abstract}
Pneumatic conveying at moderate gas velocities can result in strongly heterogeneous particle transport, with slow particles accumulating near the bottom of the pipe while faster particles are transported above. Here, we investigate helical pipes as a passive means of homogenizing particle transport without increasing the gas velocity. CFD–DEM simulations show that the helical pipe induces swirling motion of the gas and particles, continuously redistributing particles across the pipe cross-section. This transforms the heterogeneous transport observed in a straight pipe into a more homogeneous state with a narrower particle-velocity distribution. The results demonstrate that helical pipes can maintain efficient pneumatic conveying at moderate gas velocities while reducing excessive particle velocities.
\end{abstract}




\maketitle

\section{Introduction}
\label{sec:intro}
Pneumatic conveying is widely used for the transport of granular materials in industrial processes. Depending on gas velocity, solids loading, particle properties, and pipe configuration, markedly different flow behaviors can occur \cite{pan_material_1999, klinzing_historical_2018, klinzingReviewPneumaticConveying2018a, kalmanFlowRegimeChart2020a, kuangCFDmodellingSimulationPneumatic2020}. At sufficiently high gas velocities, particles are predominantly suspended and transported in dilute flow. Reducing the gas velocity lowers particle velocities and may reduce energy consumption, particle attrition, and pipe wear, but simultaneously increases the influence of gravity. In horizontal pipes, particles then increasingly accumulate near the bottom of the pipe, resulting in pronounced cross-sectional variations in particle concentration and velocity \cite{fokeerCharacterisation2004}. At still lower gas velocities and increasing solids loading, dense particle structures and eventually slug flow can develop \cite{kuangComputationalInvestigationHorizontal2008, kuangMicromechanicModelingAnalysis2011, lecreps_physical_2014, miaoNumericalSimulationDensephase2020, zhangFlowRegimesHorizontal2021b, orozovicFrameworkModellingSingle2023b, orozovicBoundariesUnstableZone2025b}. Between dilute and dense transport, horizontal pneumatic conveying can therefore be highly heterogeneous, with slow or stationary particles accumulating near the bottom while faster particles are transported above.

The mechanisms underlying these different conveying states have been studied extensively using experiments and numerical simulations. Cross-sectional particle distributions in horizontal pipes result from the competing effects of gravitational settling, fluid--particle interaction, turbulent dispersion, and particle--wall and particle--particle collisions \cite{fokeerCharacterisation2004}. Particle-scale numerical methods have provided further insight into how these mechanisms change between conveying regimes. In particular, coupled computational fluid dynamics and discrete element method (CFD--DEM) simulations can reproduce dilute, heterogeneous, and dense transport states and relate their macroscopic behavior to the forces acting on individual particles \cite{kuangMicromechanicModelingAnalysis2011, kuangCFDmodellingSimulationPneumatic2020}. These studies also illustrate a fundamental difficulty of horizontal pneumatic conveying: sufficiently large gas velocities promote particle suspension and homogeneous transport, whereas reducing the gas velocity promotes gravitational settling and increasingly heterogeneous particle motion.

A possible alternative to increasing the gas velocity is to introduce transverse motion that continuously redistributes particles across the pipe cross-section. Swirling flow has previously been explored for this purpose in pneumatic conveying. \citet{liParticleVelocity2000} experimentally showed that imposed swirl substantially changes particle velocity and concentration distributions in a horizontal pipe and can improve particle transport at low gas velocities. Swirling gas motion can also be generated passively through the pipe shape. \citet{fokeerExperimental2009} demonstrated that a three-lobed helical pipe produces a pronounced swirling gas flow and characterized its downstream decay. Pipe modifications have also been investigated in the context of particle erosion, where vortex chambers, twisted tapes, and modified wall shapes alter particle trajectories and impact conditions \cite{duarteMitigatingElbowErosion2016, duarteInnovativePipeWall2017, duarteNumericalAssessmentTwo2020, zhouStudyReducingElbow2021a}. Related effects have been observed in gravity-driven granular pipe flow, where helical inner-wall structures induce transverse particle motion, redistribute particles, and suppress localized flow structures \cite{Verbucheln2015, verbuchelnHomogenizationGranularPipe2017a}. Together, these studies show that swirl and wall-induced transverse motion can strongly modify particle transport. However, the use of a continuously helical pipe to homogenize heterogeneous particle transport in horizontal pneumatic conveying has received little attention.

In this work, we investigate a pipe with a circular cross-section whose centerline follows a helical path. Coupled CFD--DEM simulations are used to compare particle transport in straight and helical pipes over a range of gas velocities and helical configurations. The helical pipe induces swirling motion of the gas and particles, resulting in continuous transverse redistribution of the particulate phase. We show that this motion transforms the heterogeneous particle transport observed in a straight pipe at moderate gas velocities into a homogeneous conveying state. The influence of the helix amplitude and period on particle transport is examined, and the resulting transport efficiency and homogeneity are compared with those of conventional straight-pipe conveying. The results demonstrate that helical pipes provide a passive means of homogenizing pneumatic particle transport without requiring the high gas velocities associated with dilute conveying.

\section{Simulation setup and model}
\subsection{Fluid-driven flow through helical pipes}

\begin{figure}[pos=htb]
    \centering
    \includegraphics[width=1.0\columnwidth]{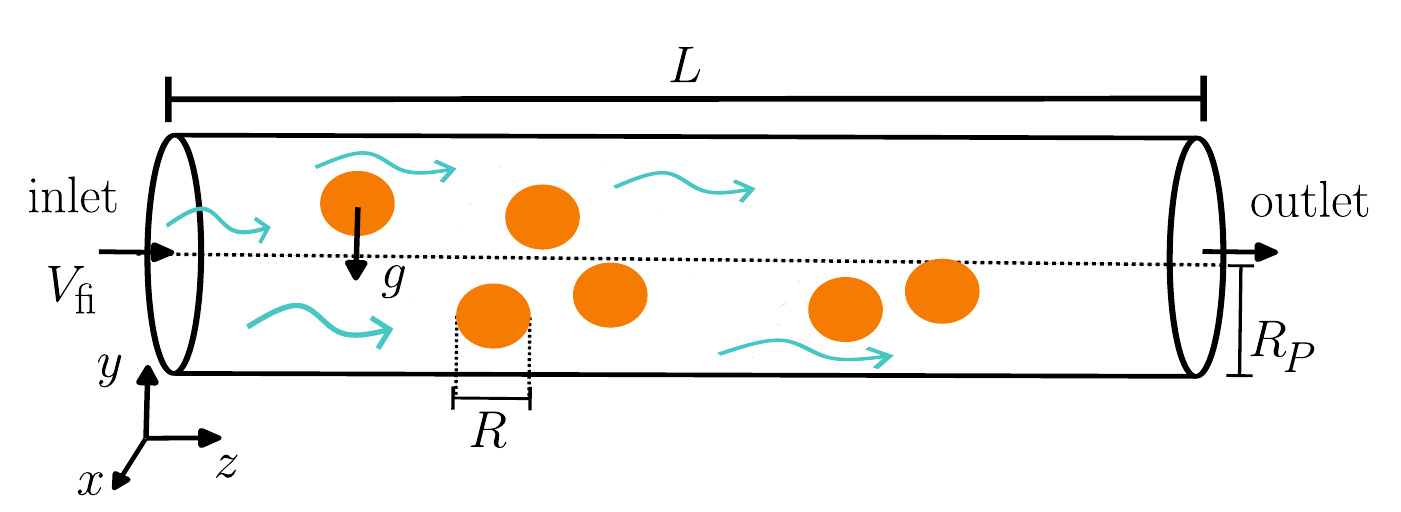}
    \includegraphics[width=1.0\columnwidth]{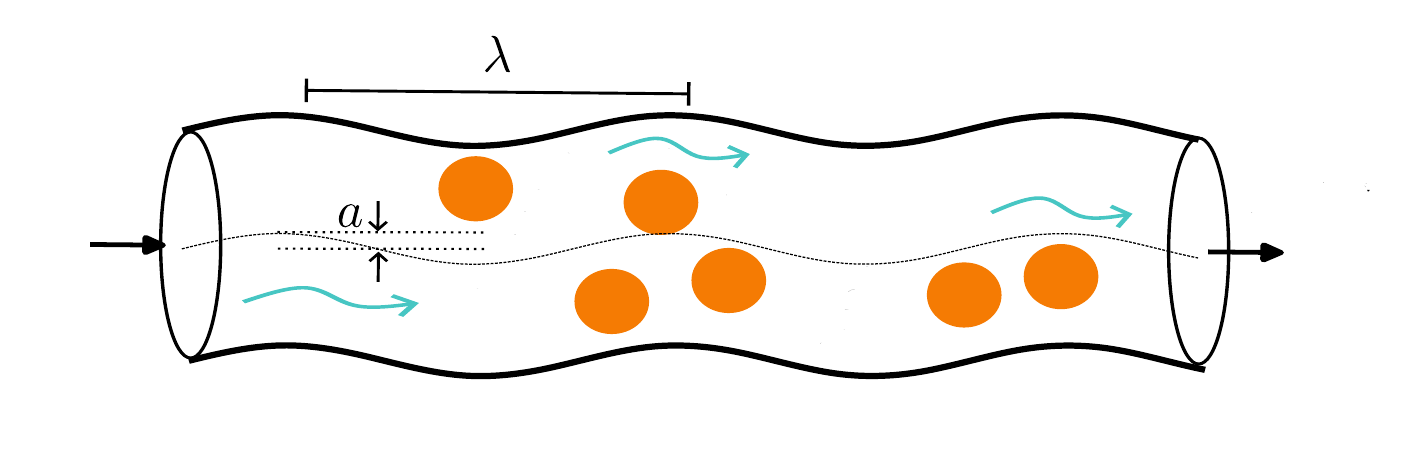}
	\caption{Simulation setup. Upper: straight pipe. Lower: helical pipe.}
    \label{fig:simSetup}
\end{figure}

We simulate fluid-driven granular flow through straight and helical pipes (\Cref{fig:simSetup}). 
Both pipes have radius $R_\text{p}$ and length $L$. The helical pipe has amplitude $a$ and period $\lambda$. The inlet and outlet are normal to the $z$-axis. Gravity acts in the negative $y$-direction.

Particles of radius $R$ enter the pipe with a random velocity in the $xy$-plane with maximum magnitude $v_0$. Air enters with velocity $V_\text{fi}$ in the positive $z$-direction. Particles are removed at the outlet.

\subsection{Unresolved fluid-particle coupling}

We use the computational fluid dynamics--discrete element method (CFD--DEM) \cite{gallas1992SimA,gallas1992SimB,Poschel2005,kloss_models_2012}, implemented in OpenFOAM \cite{weller_tensorial_1998} and LIGGGHTS \cite{kloss_models_2012}. The fluid motion is solved on a computational mesh that discretizes the pipe. The particles move independently through this mesh and are tracked individually by the DEM.

We use an unresolved CFD--DEM approach, in which the fluid mesh does not resolve the flow around individual particles \cite{shirgaonkar_new_2009}. Instead, the presence of the particles is represented by the local fluid volume fraction. For a fluid cell of volume $V_\text{cell}$,
\begin{equation}
\alpha_\text{f}
=
1-\frac{\sum_{i\in\text{cell}} V_i}{V_\text{cell}},
\end{equation}
where $V_i$ is the volume of particle $i$ contained in the cell. Fluid properties and fluid--particle interaction forces are evaluated on the fluid mesh and exchanged between the CFD and DEM solvers.

The fluid phase is governed by the volume-averaged Navier--Stokes equations\cite{ZHOU_KUANG_CHU_YU_2010,GONIVA2012582}. Conservation of mass gives
\begin{equation}
\frac{\partial\alpha_\text{f}}{\partial t}
+
\nabla\cdot
\left(
\alpha_\text{f}\,\vec{u}_\text{f}
\right)
=0,
\end{equation}
and conservation of momentum gives
\begin{equation}
\frac{\partial
\left(
\alpha_\text{f}\,\vec{u}_\text{f}
\right)}
{\partial t}
+
\nabla\cdot
\left(
\alpha_\text{f}\,
\vec{u}_\text{f}\,\vec{u}_\text{f}
\right)
=
-\frac{\alpha_\text{f}}{\rho_\text{f}}\,\nabla P
+
\nabla\cdot
\left(
\alpha_\text{f}\,\tau_\text{f}
\right)
+
\vec{R}_\text{pf}\,.
\end{equation}
Here, $\rho_\text{f}$, $\vec{u}_\text{f}$, and $P$ are the fluid density, velocity, and pressure. The kinematic viscous stress tensor is
\begin{equation}
    \tau_\text{f}
=
\nu_\text{f}
\left[
    \nabla\vec{u}_\text{f}
    +
    \left(\nabla\vec{u}_\text{f}\right)^T
    -
    \frac{2}{3}
    \left(\nabla\cdot\vec{u}_\text{f}\right)I
\right]\,.
\end{equation}
where $\nu_\text{f}$ is the kinematic viscosity.

The momentum transferred from the particles to the fluid is represented by the source term
\begin{equation}
\vec{R}_\text{pf}
=
-\frac{
\left|
\sum_{i\in\text{cell}} \vec{F}_{\text{drag},i}
\right|
}{
\rho_\text{f}V_\text{cell}
}
\frac{
\vec{u}_\text{f}-\left<\vec{u}\right>_\text{p}
}{
\left|
\vec{u}_\text{f}-\left< \vec{u}\right>_\text{p}
\right|
}\,,
\end{equation}

where $\left< \vec{u} \right>_\text{p}$ is the mean particle velocity in the cell and $\vec{F}_{\text{drag},i}$ is the fluid drag force on particle $i$, that will be described in \cref{sec:fluid-force}.

A uniform fluid velocity $V_\text{fi}$ in the positive $z$-direction is prescribed at the inlet. The outlet has zero pressure gradient, and a no-slip condition is imposed at the pipe wall. The motion of each particle is obtained from Newton's equation of motion,
\begin{equation}
m_\text{p}\frac{d\vec{u}_\text{p,i}}{dt}
=
m_\text{p} \vec{g}
+
\vec{F}_{\text{pf,i}}
+
\vec{F}_{\text{c}}\,,
\end{equation}
where  $\vec{u}_\text{p,i}$ is the velocity of particle i, $m_\text{p}$ is its mass, $\vec{F}_{\text{pf,i}}$ is the fluid--particle interaction force, and $\vec{F}_{\text{c}}$ stands for the particle--particle and particle--wall contact forces. Contacts are modeled by a viscoelastic Hertzian normal force \cite{brilliantovModelCollisionsGranular1996} and a modified Cundall--Strack tangential force \cite{cundallDiscreteNumericalModel1979, kloss_models_2012}. A full description of the DEM model containing all details and motivations is given in \cite{mullerParticleSizeSegregation2025}.

\subsection{Fluid--particle interaction forces}
\label{sec:fluid-force}

The interaction force exerted by the fluid on the particle i consists of the pressure-gradient force, viscous force and drag force 
\begin{equation}
\vec{F}_{\text{pf,i}}
=
\vec{F}_{\text{pressure,i}}
+
\vec{F}_{\text{viscous,i}}
+
\vec{F}_{\text{drag,i}}
\,.
\end{equation}
The pressure-gradient force and the viscous force are
\begin{equation}
\vec{F}_{\text{pressure,i}}
=
-V_\text{p} \nabla P\,,
\end{equation}
\begin{equation}
\vec{F}_{\text{viscous,i}}
=
-V_\text{p} \nabla \cdot (\rho_f\tau_f)\,,
\end{equation}
where $V_\text{p}$ is the particle volume.

The drag force is calculated using the Koch--Hill model \cite{hill_moderate-reynolds-number_2001},
\begin{equation}
\vec{F}_{\text{drag,i}}
=
V_\text{p}
\frac{9\mu_\text{f} \, \alpha_\text{f}^2}{2R^2}
\left[
F_0\left(\alpha_\text{p}\right)
+
\frac{1}{2}F_3 \left(\alpha_\text{p}\right) Re_\text{p,i}
\right]
\left(
\vec{u}_\text{f}-\vec{u}_\text{p,i} \right)\,,
\end{equation}
where $\mu_\text{f}$ is the dynamic viscosity and $\alpha_\text{p}=1-\alpha_\text{f}$ is the local particle volume fraction. The particle Reynolds number is
\begin{equation}
Re_\text{p,i}
=
\frac{
2\alpha_\text{f}\rho_\text{f}
\left|
\vec{u}_\text{f}-\vec{u}_\text{p,i}
\right|
R
}{
\mu_\text{f}
}\,.
\end{equation}
$F_0$ and $F_3$ are functions of the local particle volume fraction,
\begin{equation}
F_0\left(\alpha_\text{p}\right) =
\begin{cases}
\displaystyle
\frac{
1+3\sqrt{\frac{\alpha_\text{p}}{2}}
+\frac{135}{64}\alpha_\text{p}\ln \alpha_\text{p}
+16.14\alpha_\text{p}
}{
1+0.681\alpha_\text{p}
-8.48\alpha_\text{p}^2
+8.16\alpha_\text{p}^3
}
&
\!\!\!,\,\alpha_\text{p}\leq0.4
\\[3mm]
\displaystyle
\frac{10\alpha_\text{p}}
{(1-\alpha_\text{p})^3}
&
\!\!\!,\,\alpha_\text{p}>0.4
\end{cases}
\end{equation}
\begin{equation}
F_3\left(\alpha_\text{p}\right)
=
0.0673
+
0.212\,\alpha_\text{p}
+
\frac{0.0232}
{\left(1-\alpha_\text{p}\right)^5}\,.
\end{equation}

\subsection{Simulation parameters}

The material and system parameters are summarized in \Cref{table:param}. 
\begin{table}[htbp]
\caption{\label{table:param}Simulation parameters. 
} 
\centering
\begin{tabular}{|l|l|l|}
\hline
parameter & & value \\
\hline
particle material density & $\rho_\text{p}$ & $890\,\text{kg/m}^3$ \\
Young modulus & $E$ & $10\,\mathrm{MPa}$ \\
Poisson ratio & $\nu$ & $0.24$ \\
friction coefficient & $\mu$ & $0.5$ \\
fluid density & $\rho_\text{f}$ & $1\,\mathrm{kg/m^3}$ \\
kinematic viscosity & $\nu_\text{f}$ & $1.53\times10^{-5}\,\mathrm{m^2/s}$ \\
particle radius & $R$ & $1.59\,\mathrm{mm}$ \\
pipe radius & $R_\text{p}$ & $28.62\,\mathrm{mm}$ \\
pipe length & $L$ & $2000\,\mathrm{mm}$ \\
maximum initial particle velocity & $v_0$ & $0.1\,\mathrm{m/s}$ \\
inlet fluid velocity & $V_\text{fi}$ & $5$--$20\,\mathrm{m/s}$ \\[0.5ex]
\hline
\end{tabular}
\end{table}

The fluid mesh has a minimum cell size of $dx=5\,\mathrm{mm}$; the corresponding cell volume is approximately nine times the particle volume. This mesh size limits small local values of $\alpha_\text{f}$ while retaining sufficient spatial resolution of the fluid flow. The time step is chosen according to the Courant--Friedrichs--Lewy (CFL) criterion \cite{courantPartialDifferenceEquations1967} and the Rayleigh criterion for the DEM \cite{burnsCriticalTimeStep2019}. The CFL limit is
\begin{equation}
T_\text{CFL}
=
\frac{C_\text{max}dx}{U_\text{max}}
=
2.5\times10^{-5}\,\mathrm{s}\,,
\end{equation}
where $C_\text{max}=0.1$ and $U_\text{max}=20\,\mathrm{m/s}$ is the maximum inlet fluid velocity. The Rayleigh limit is
\begin{equation}
T_\text{R}
=
\frac{\pi R}
{0.1631\nu+0.8766}
\sqrt{
\frac{2(1+\nu)\rho_\text{p}}{E}
}
=
8.10\times10^{-5}\,\mathrm{s}\,.
\end{equation}
In our simulations, we use the time step 
\begin{equation}
T_\text{step}=10^{-6}\,\mathrm{s}\,,
\end{equation}
which fulfills both criteria and, thus, ensures stability for the coupled solver.

Particles are continuously supplied at the inlet by maintaining a particle volume fraction close to unity in the inlet region. The particle mass flow rate is therefore not prescribed. Instead, it results from the rate at which the fluid entrains and transports particles through the pipe.
Each simulation is run until the particle flow reaches a statistically stationary state. Only data collected after this state is reached are used for the analysis.

\subsection{Flow characterization}
We characterize particle transport by the mean axial particle velocity, the maximum particle velocity, and the Gini coefficient of the axial particle velocity \cite{biro_gintropy_2020, dingwall_large_2023}. The mean axial velocity measures the transport velocity, while the maximum velocity identifies particles with particularly high velocities. The Gini coefficient measures the heterogeneity of the particles' velocities. All quantities are evaluated for particles in the downstream half of the pipe, $z>L/2$, where the influence of the inlet conditions has decayed and the particle flow represents the established conveying state. 

The instantaneous mean axial particle velocity is
\begin{equation}
\left<{v}_z\right>
=
\frac{1}{n}
\sum_{i=1}^{n}v_{z,i},
\end{equation}
where $n$ is the number of particles in the evaluation region and $v_{z,i}$ is the axial velocity of particle $i$. The maximum particle velocity is then
\begin{equation}
v_\text{max}
=
\max_{i}
\left|\vec{u}_i\right|\,.
\end{equation}
(Note that $v_\text{max}$ is the maximal value of the velocity, but not the maximal velocity along the pipe direction, $z$.)

Both quantities are then averaged over time after the flow has reached a statistically stationary state.

The heterogeneity of the axial particle velocities is quantified by the Gini coefficient,
\begin{equation}
G
\equiv
\frac{
\displaystyle
\sum_{i=1}^{n}
\sum_{j=1}^{n}
\left|
v_{z,i}-v_{z,j}
\right|
}{
2n^2 \left< v_z \right>
}\,.
\end{equation}
$G=0$ corresponds to identical axial velocities of all particles. Larger values of $G$ indicate a broader and more unequal velocity distribution. The Gini coefficient, therefore, distinguishes homogeneous transport from states in which slow or stationary particles coexist with rapidly transported particles. The reported Gini coefficient is averaged over time after the statistically stationary state is reached.

\section{Results and discussion}
\subsection{Particle transport in the straight pipe}
\begin{figure}[pos=htb]
        \centering
        \includegraphics[width=\columnwidth]{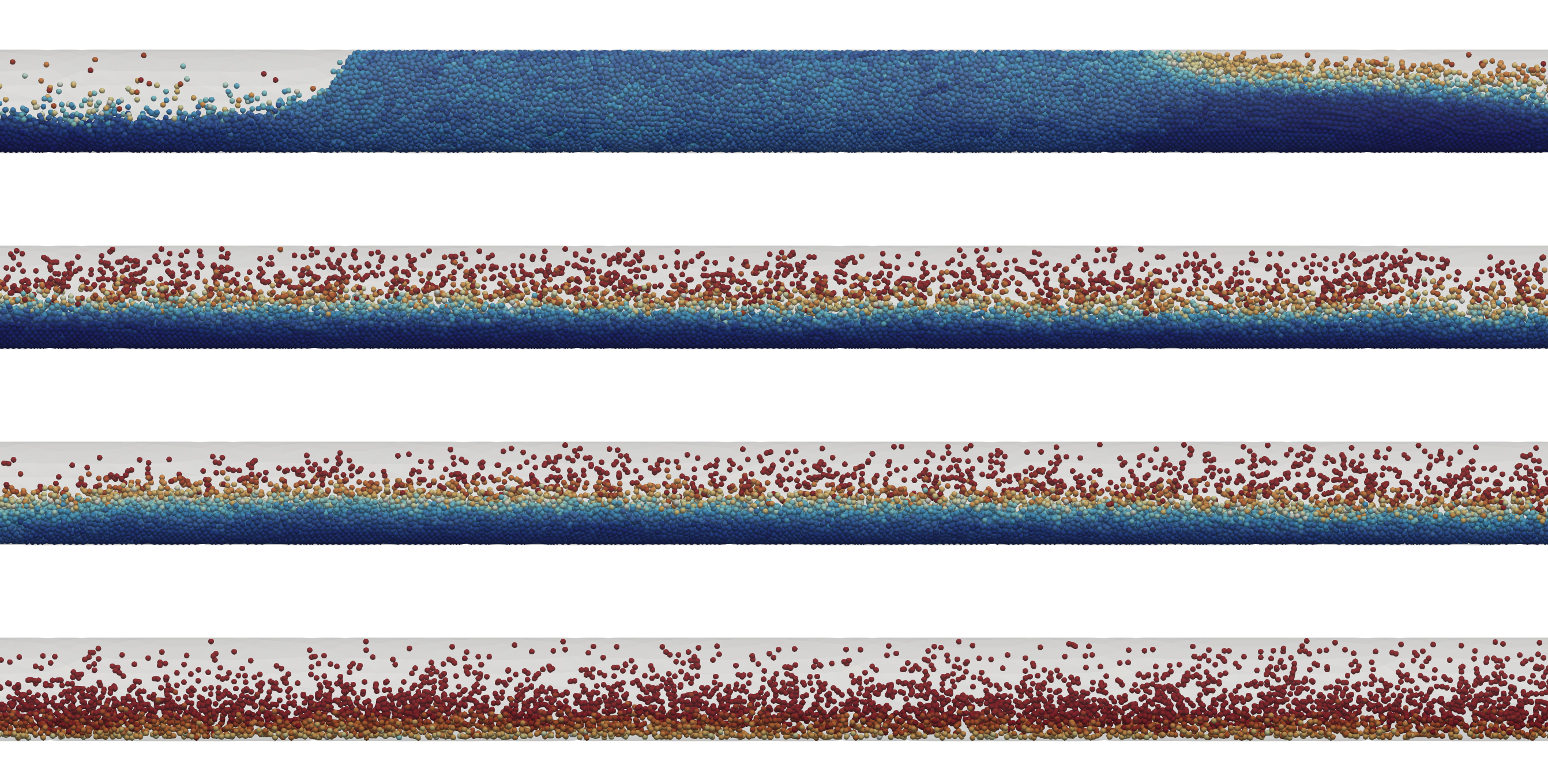}
    \caption{Particle flow in the straight pipe at different inlet fluid velocities. Color indicates particle velocity, increasing from blue to red. From top to bottom: $V_\text{fi}=5$, $10$, $15$, and $20\,\mathrm{m/s}$.}
    \label{fig:visualflowreference}
\end{figure}

We first examine particle transport in the straight reference pipe for inlet fluid velocities $V_\text{fi}=5$, $10$, $15$, and $20\,\mathrm{m/s}$. \Cref{fig:visualflowreference} shows the resulting particle distributions and velocities.

At $V_\text{fi}=5\,\mathrm{m/s}$, the fluid cannot continuously transport the supplied particles. Particles accumulate and form dense structures that intermittently move through the pipe, resulting in slug flow.

Increasing the fluid velocity to $V_\text{fi}=10\,\mathrm{m/s}$ prevents the formation of persistent slugs, but does not produce homogeneous particle transport. Most particles remain close to the bottom of the pipe and move slowly, while a smaller fraction is transported rapidly above them. The particle flow is therefore strongly heterogeneous.

This heterogeneity decreases with increasing fluid velocity. At $V_\text{fi}=15\,\mathrm{m/s}$, more particles are suspended and transported along the pipe, although slow particles still accumulate near the bottom. At $V_\text{fi}=20\,\mathrm{m/s}$, the particles are predominantly suspended and transported in dilute flow.

\begin{figure}[pos=htb]
\centering
\includegraphics[width=1.0\columnwidth]{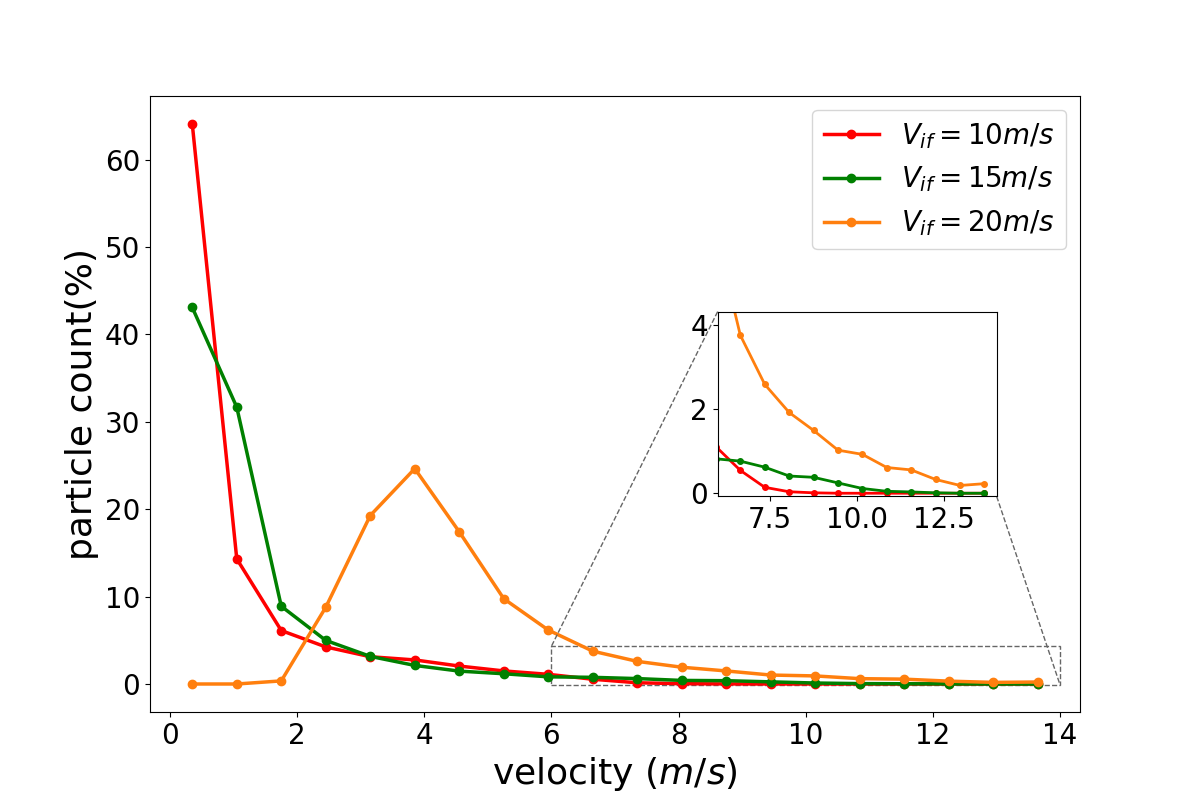}
	\caption{Particle velocity distributions in the straight pipe at different inlet fluid velocities.
    }
    \label{fig:referencePipeDistribution}
\end{figure}

The corresponding axial particle-velocity distributions are shown in \Cref{fig:referencePipeDistribution}. 
For $V_\text{fi}\leq15\,\mathrm{m/s}$, the distributions contain a large fraction of particles with velocities close to zero and a smaller fraction with much higher velocities. The broad distributions reflect the coexistence of slowly moving particles near the bottom and rapidly transported particles above them.

At $V_\text{fi}=20\,\mathrm{m/s}$, the distribution changes markedly. Most particles move at approximately $4\,\mathrm{m/s}$, and the fraction of nearly stationary particles is strongly reduced. However, the distribution retains a high-velocity tail, with some particles moving considerably faster than the majority. Increasing the fluid velocity therefore improves particle suspension and reduces velocity heterogeneity, but also increases the velocities reached by the fastest particles.

\subsection{Particle transport in the helical pipe}
\begin{figure}[pos=htb]
        \centering
        \includegraphics[width=\columnwidth]{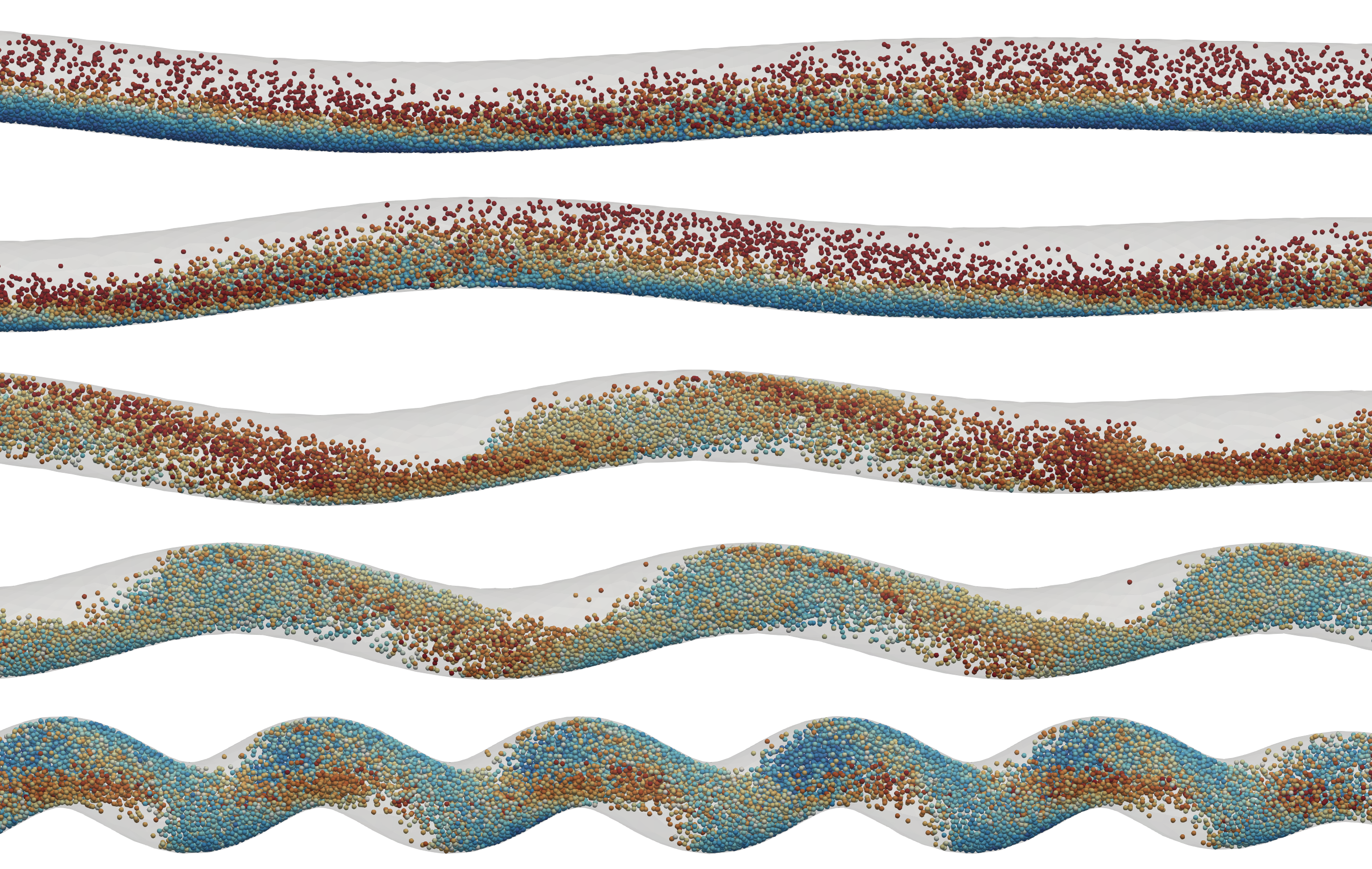}
    \caption{Particle flow in the helical pipe at different helix periods for $a=9R$ and $V_\text{fi}=15\,\mathrm{m/s}$. Color indicates particle velocity, increasing from blue to red. From top to bottom: $\lambda=500\,R$, $400\,R$, $300\,R$, $200\,R$, and $100\,R$.}
    \label{fig:visualflow}
\end{figure}

We now consider the helical pipe at $V_\text{fi}=15\,\mathrm{m/s}$, where particle transport in the straight pipe is strongly heterogeneous. \Cref{fig:visualflow} 
shows the particle flow for a fixed helix amplitude $a=9R$ and different helix periods $\lambda$.

For large $\lambda$, the particle flow remains similar to that in the straight pipe. Slow particles accumulate near the bottom, while faster particles are transported above them. Decreasing $\lambda$ increasingly redistributes the particles across the pipe cross-section. At $\lambda=200R$, particles are distributed throughout the pipe and the pronounced separation between slow and fast particles largely disappears.

A further decrease to $\lambda=100R$ does not improve the homogenization. Instead, a velocity difference develops between particles near the wall and particles closer to the pipe center. The effect of the helical pipe is therefore non-monotonic: decreasing the helix period initially homogenizes particle transport, but a very short period can increase velocity heterogeneity again.

\begin{figure}[pos=htb]
        \centering
        \includegraphics[width=\columnwidth]{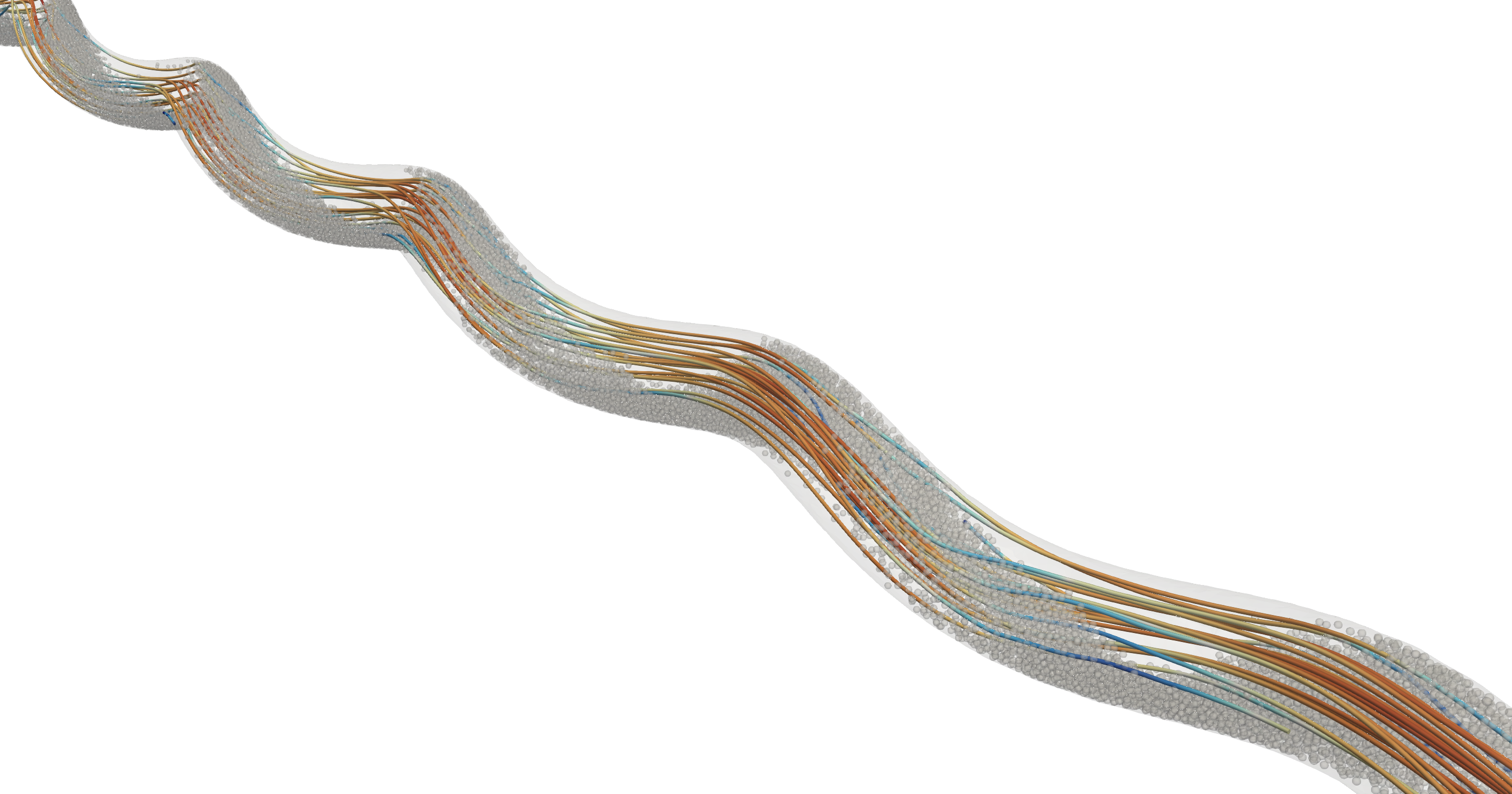}
    \caption{Fluid streamlines in the helical pipe for $\lambda=200 R$ and $a=9 R$. Color indicates fluid speed, increasing from blue to red. The fluid preferentially follows the inner bend, while the particles are concentrated near the outer bend. Particles are shown transparently to improve the visibility of the streamlines.}
    \label{fig:visualflowstreamlines}
\end{figure}

The redistribution of the particles results from the transverse fluid motion induced by the helical pipe. \Cref{fig:visualflowstreamlines} 
shows the fluid streamlines for $a=9R$ and $\lambda=200R$. In addition to the axial motion, the fluid develops a pronounced swirling motion around the pipe axis. This transverse motion carries particles around the pipe cross-section and counteracts their persistent accumulation at the bottom. The particles consequently follow helical trajectories through the pipe rather than remaining confined to a lower particle-rich region.

\subsection{Effect of the helix parameters}
\begin{figure}[pos=htb]
\centering
\includegraphics[width=1.0\columnwidth]{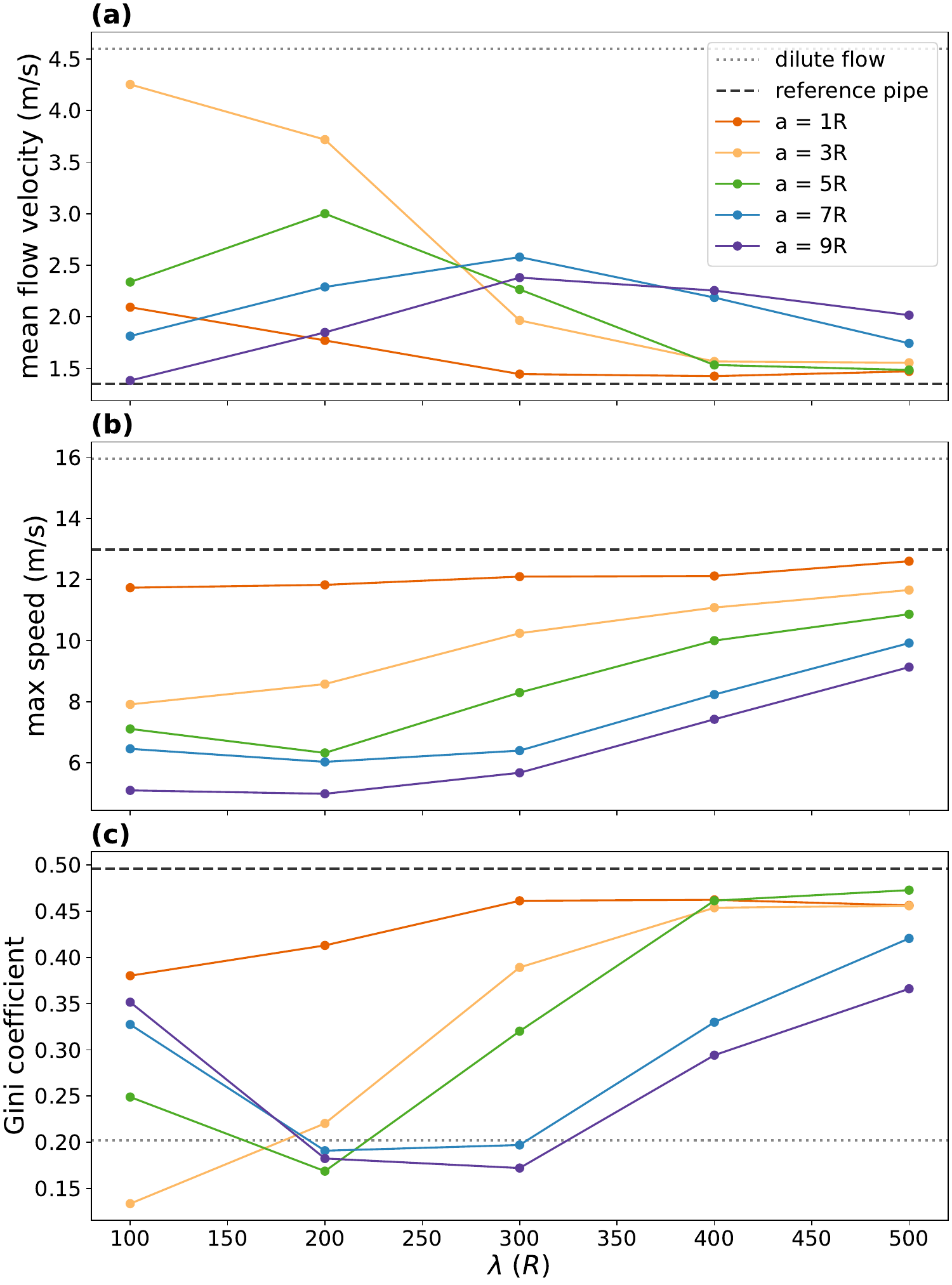}
\caption{Mean axial particle velocity, maximum particle speed, and Gini coefficient for different helix amplitudes $a$ and periods $\lambda$ at $V_\text{fi}=15\,\mathrm{m/s}$. Results for the straight pipe at $V_\text{fi}=15\,\mathrm{m/s}$ and dilute flow at $V_\text{fi}=20\,\mathrm{m/s}$ are shown for comparison.
}
\label{fig:comparehelix}
\end{figure}

We next quantify the influence of the helix amplitude $a$ and period $\lambda$ on particle transport. \Cref{fig:comparehelix} 
shows the mean axial particle velocity, maximum particle speed, and Gini coefficient for the different helical pipes at $V_\text{fi}=15\,\mathrm{m/s}$. The straight pipe at the same inlet velocity is used as the reference. The dilute flow in the straight pipe at $V_\text{fi}=20\,\mathrm{m/s}$ is included for comparison.

The Gini coefficient shows that the degree of homogenization depends strongly on both $a$ and $\lambda$. For amplitudes up to $a=3R$, $G$ decreases with decreasing $\lambda$. For $a\geq5R$, decreasing $\lambda$ initially also improves homogenization, but this trend reverses for $\lambda<200R$. 
For large amplitudes, a further decrease of the helix period below $200 R$ does not improve homogenization and instead increases the Gini coefficient.
This agrees with \Cref{fig:visualflow}, where slow particles accumulate near the outer part of the pipe for $a=9 R$ and $\lambda=100 R$.

The lowest Gini coefficient, $G\approx 0.12$, is obtained for $a=3 R$ and $\lambda=100 R$. The particle velocities in this configuration are more homogeneous than in the dilute straight-pipe flow at $V_\text{fi}=20\,\mathrm{m/s}$. Thus, increasing the gas velocity is not required to obtain a narrow particle-velocity distribution.

The helical pipe also changes the particle transport velocity. The mean axial particle velocity is higher than in the straight pipe at $V_\text{fi}=15\,\mathrm{m/s}$ for all investigated helical configurations. The largest mean velocity, approximately $4.2\,\mathrm{m/s}$, occurs for $a=3 R$ and $\lambda=100 R$, which also gives the lowest Gini coefficient. The homogenization is therefore accompanied by faster particle transport rather than by a reduction in the mean particle velocity.

At the same time, the maximum particle speed remains below that of the dilute straight-pipe flow. The helical pipe therefore combines a high mean transport velocity with a narrower range of particle velocities. This distinction is important because the same mean velocity can result from markedly different particle-velocity distributions.

\begin{figure}[pos=htb]
\centering
\includegraphics[width=1.0\columnwidth]{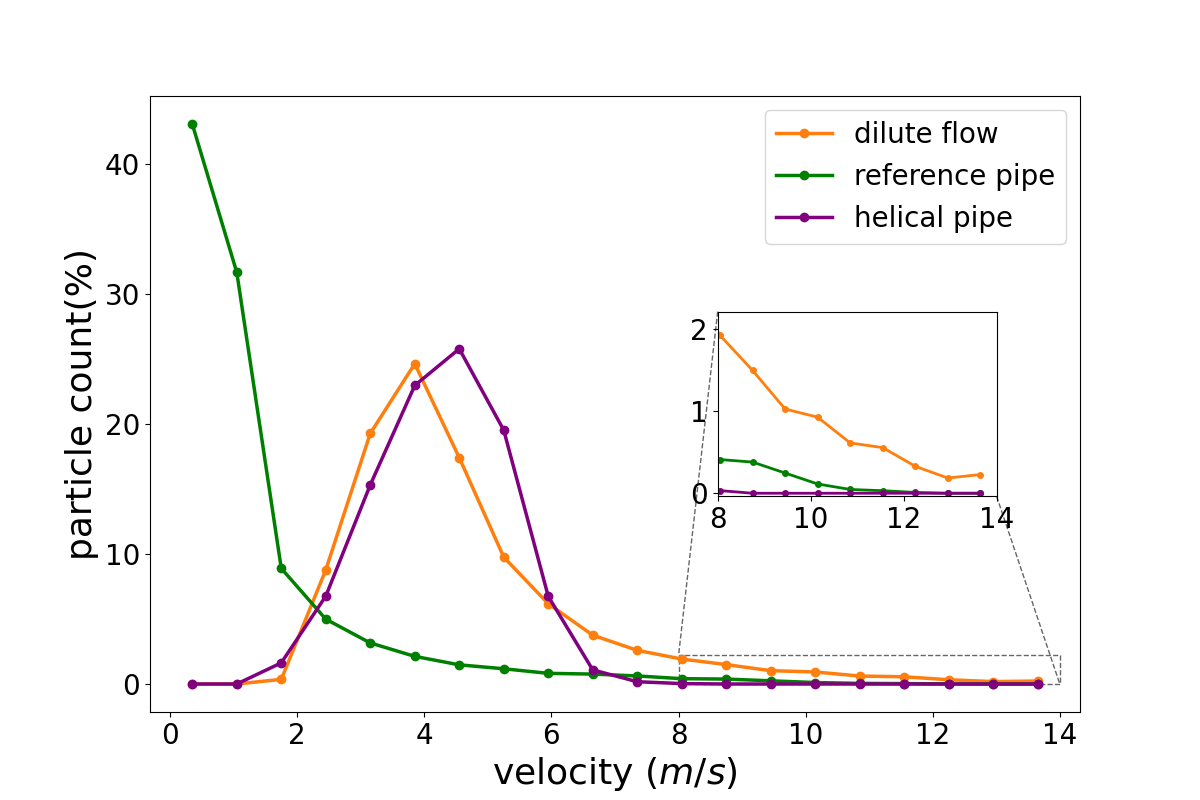}
\caption{Particle velocity distributions in the helical pipe with $a=3R$ and $\lambda=100R$ at $V_\text{fi}=15\,\mathrm{m/s}$ and in the straight pipe at different inlet fluid velocities.
}
\label{fig:helicalPipeDistribution}
\end{figure}

\Cref{fig:helicalPipeDistribution} compares the particle velocity distribution for the helical pipe with $a=3R$ and $\lambda=100R$ with those in the straight pipe.
At $V_\text{fi}=15\,\mathrm{m/s}$, the distributions in the straight and helical pipes differ substantially. In the straight pipe, a large fraction of the particles moves slowly, while a smaller fraction reaches much higher velocities. In the helical pipe, the fraction of slow particles is strongly reduced and the distribution becomes considerably narrower. This is consistent with the low Gini coefficient in \Cref{fig:comparehelix}.

The mean particle velocity in the helical pipe at $V_\text{fi}=15\,\mathrm{m/s}$ is comparable to that of the dilute straight-pipe flow at $V_\text{fi}=20\,\mathrm{m/s}$. Their velocity distributions, however, are different. The dilute straight-pipe flow retains a high-velocity tail, whereas the helical pipe concentrates the particle velocities in a narrower range. The helical pipe therefore achieves a similar mean transport velocity at a lower inlet fluid velocity while reducing the velocities reached by the fastest particles.

These results show that the principal effect of the helical pipe is not simply an increase in particle velocity. The transverse particle motion reduces the coexistence of slow particles near the bottom and rapidly transported particles above them. The resulting homogenization shifts particle transport from a broad velocity distribution toward a state in which a larger fraction of the particles contributes to the axial transport at comparable velocities.

\section{Conclusion}
\label{sec:conclusion}

We investigated particle transport through continuously helical pipes using coupled CFD--DEM simulations. At moderate fluid velocities, particle transport in a straight horizontal pipe is strongly heterogeneous, with slow particles accumulating near the bottom while faster particles are transported above. The helical pipe induces transverse fluid and particle motion that continuously redistributes particles across the pipe cross-section and suppresses this separation.
The degree of homogenization depends on the helix amplitude and period. Decreasing the helix period generally improves homogenization, but very short periods combined with large amplitudes produce a new separation between slow particles near the outer part of the pipe and faster particles closer to the center. Among the investigated configurations, $a=3R$ and $\lambda=100R$ produces the lowest Gini coefficient, $G\approx0.12$.
For this configuration at $V_\text{fi}=15\,\mathrm{m/s}$, the mean particle velocity is comparable to dilute transport in the straight pipe at $V_\text{fi}=20\,\mathrm{m/s}$, while the particle-velocity distribution is narrower and the maximum particle speed is lower. The helical pipe therefore enables homogeneous particle transport at a lower fluid velocity without sacrificing the mean particle transport velocity.
The lower maximum particle velocities may reduce particle attrition and impact-induced wear, but these effects were not evaluated directly. Likewise, a lower inlet fluid velocity does not necessarily imply lower energy consumption because the pressure drop through the helical pipe must also be considered. The present results establish that helical pipes provide a passive means of converting heterogeneous particle transport into homogeneous transport while avoiding the high particle velocities associated with dilute conveying in a straight pipe.

\printcredits
\appendix

\bibliographystyle{cas-model2-names}

\bibliography{fluidpipe}

\end{document}